\documentclass[twocolumn,abs,prb,showpacs]{revtex4-1}

\usepackage{graphicx}
\usepackage{dcolumn}
\usepackage{float}
\usepackage{amsmath}
\usepackage{bm}

\newcommand{\comment}[1]{}

\begin{document}

%\preprint{}

\title{Dynamical Polarization Function, Plasmons and Screening in Silicene and other Buckled Honeycomb Lattices}

\author{C.J. Tabert}
\author{E.J. Nicol}
\affiliation{Department of Physics, University of Guelph,
Guelph, Ontario N1G 2W1 Canada} 
\affiliation{Guelph-Waterloo Physics Institute, University of Guelph, Guelph, Ontario N1G 2W1 Canada}
\date{\today}

\begin{abstract}
{ We explore the dielectric properties of graphene-like two-dimensional Kane-Mele topological insulators manifest in buckled honeycomb lattices (ex. silicene, germanene, etc.).  The effect of an onsite potential difference ($\Delta_z$) between sublattices is given particular attention.  We present the results for the real and imaginary parts of the dynamical polarization function.  We show that these results display features of three regimes (topological insulator, valley-spin polarized metal, and trivial band insulator) and may be used to extract information on the strength of the intrinsic spin-orbit coupling.  We study the inverse dielectric function and provide numerical results for the plasmon branch.  We discuss the behaviour of the plasmon as a function of sublattice potential difference and show that the behaviour of the plasmon branch as $\Delta_z$ is varied is dependent on the location of the chemical potential with respect to the gaps.  The static polarization is discussed and numerical results for the screening of a charged impurity are provided.  We observe a beating phenomenon in the effective potential which is dependent on $\Delta_z$.
}
\end{abstract}

 %dielectric function/plasmons, permittivity(dielectric function), graphene

\maketitle
% body of paper here

\section{Introduction}
Graphene has attracted considerable attention since it was successfully isolated in 2004.  Many of graphene's novel properties stem from its low-energy band structure.  At zero energy, the valence and conduction bands of graphene meet each other at two inequivalent $K$ points, $\bm{K}$ and $\bm{K}^\prime=-\bm{K}$, in the first Brillouin zone. At these points, the energy spectrum is well described by a simple linear relation $E_{\bm{k}}=\pm\hbar v_F|\bm{k}|$ where $v_F\approx 10^6$m/s is the Fermi velocity and $|\bm{k}|$ is the momentum measured relative to the $K$ point.  While this feature promises many exciting possibilities, some of graphene's potential is limited by the difficulty of opening a sizable insulating gap in the band structure.  As a result, attention has begun to shift to other two-dimensional (2D) systems such as graphene bilayers and systems made of different elements.  

A promising class of 2D crystals are those with sizable intrinsic spin-orbit coupling, the low-energy Hamiltonian of which maps onto a Kane-Mele type Hamiltonian\cite{Kane:2005}.  These systems have a spin-orbit induced band gap of $2\Delta_{\rm so}$.  Examples of such crystals include silicene ($2\Delta_{\rm so}\approx 1.55-7.9$meV\cite{Liu:2011, Liu:2011a, Drummond:2012}) and germanene ($2\Delta_{\rm so}\approx 24-93$meV\cite{Liu:2011, Liu:2011a}).  These lattices are based on the 2D honeycomb of graphene but have an out of plane buckling such that the $A$ and $B$ sublattices sit in vertical planes separated by a distance $d$ ($\approx 0.46$\AA\, for silicene\cite{Drummond:2012,Ni:2012}). This asymmetry causes an onsite potential difference of $\Delta_z=E_zd$ to occur between sublattices when an out-of-plane electric field is applied.  This allows for a tuning of the band gap and the bands become spin split.  Among other interesting features, this spin-splitting has been predicted to allow for optically-generated spin-valley-polarized charge carriers\cite{Stille:2012, Tabert:2013a, Tabert:2013b, Tabert:2013c}.  It has been theoretically argued\cite{Drummond:2012, Ezawa:2012a} that as $\Delta_z$ is increased to a value greater than $\Delta_{\rm so}$, the system transitions from a topological insulator (TI) to a band insulator (BI).  At the critical value $\Delta_z=\Delta_{\rm so}$, the lowest band gap closes and the system is referred to as a valley-spin polarized metal (VSPM)\cite{Ezawa:2012}.

In the study of graphene, the dynamical polarization function $\Pi(\omega,q)$ is one of the fundamental quantities required to understand physical properties of the system such as screening due to a charged impurity and collective excitations (such as plasmons).  $\Pi(\omega,q)$ has been extensively  studied for graphene both with\cite{Pyat:2008, Pyat:2009, Scholz:2012} and without\cite{Shung:1986, Gorbar:2002, Ando:2006, Wunsch:2006, Barlas:2007, Hwang:2007, Kotov:2012, Stauber:2013, Roldan:2013} an energy gap.  In this paper, we examine the dynamical polarization function for buckled honeycomb lattices which exhibit an intrinsic spin-orbit coupling and a band gap that can be tuned by changing the onsite potential difference between sublattices.  

A particular quantity of interest in 2D systems is the plasmon.  Plasmons are collective excitations of oscillating charge.  In the system discussed here, we show that the location of the plasmon branch in $(q,\omega)$ space is dependent on the sublattice potential difference.  Recently, there has been a lot of experimental progress\cite{Luo:2013} in observing this phenomenon in graphene and similar techniques should apply to the systems discussed here.

Our paper is organized as follows: In Sec.~II we provide a brief outline of the low-energy model that was used along with a short discussion of the resulting band structure.  In Sec.~III we provide analytic expressions for the dynamic and static polarization function. Section~IV contains the resulting plasmon dispersion.  The screening of a charge impurity is shown in Sec.~V and our conclusions follow in Sec.~VI.
  
\section{Low-Energy Model}

It has been shown\cite{Liu:2011a, Ezawa:2012, Ezawa:2012a, Ezawa:2012b} that the low-energy Hamiltonian of a buckled honeycomb lattice, with an intrinsic spin-orbit band gap of $2\Delta_{\rm  so}$, in the presence of a perpendicular electric field can be expressed as
\begin{equation}\label{Hamiltonian}
\hat{H}=\hbar v(\xi k_x\hat{\tau}_x+k_y\hat{\tau}_y)-\xi\Delta_{\rm so}\sigma_z\tau_z+\Delta_z\tau_z,
\end{equation} %check signs etc.
where $\tau_i$ and $\sigma_i$ are Pauli matrices associated with the pseudospin and real spin of the system, respectively. $\xi=\pm 1$ indexes over the two inequivalent $K$ points $K$ and $K^\prime$, respectively. $v$ is the Fermi velocity ($\approx 5\times 10^5$m/s for silicene\cite{Ezawa:2012a}) and $k_x$ and $k_y$ are the momentum components measured relative to the $K$ points.  The first term is the low-energy graphene Hamiltonian\cite{Neto:2009, Abergel:2010}. The second term describes a Kane-Mele system\cite{Kane:2005} for intrinsic spin-orbit coupling with an associated spin-orbit band gap of $2\Delta_{\rm so}$.  The final term describes the sublattice potential difference that could arise from the application of a perpendicular electric field\cite{Ezawa:2012, Ezawa:2012a, Ezawa:2012b, Drummond:2012}. As a matrix, Eqn.~\eqref{Hamiltonian} is block diagonal in 2x2 matrices labelled by valley ($\xi$) and spin ($\sigma=\pm 1$ for up and down spin, respectively).  These 2x2 matrices are
\begin{equation}\label{Ham}
\hat{H}_{\sigma\xi}=\left(\begin{array}{cc}
-\sigma\xi\Delta_{\rm so}+\Delta_z & \hbar v(\xi k_x-ik_y)\\
\hbar v(\xi k_x+ik_y) & \sigma\xi\Delta_{\rm so}-\Delta_z
\end{array}\right).
\end{equation}

This yields the low-energy eigenvalues
\begin{equation}
\pm E_{\bm{k}}=\pm\sqrt{\hbar^2v^2|\bm{k}|^2+\Delta_{\xi\sigma}^2},
\end{equation}
where $\Delta_{\sigma\xi}=|\sigma\xi\Delta_{\rm so}-\Delta_z|$.  When discussing results, it is convenient to define $\Delta_{\rm max}\equiv\Delta_{+-/-+}$ and $\Delta_{\rm min}\equiv\Delta_{++/--}$ to correspond to the maximum and minimum gaps at each valley, respectively.  
 \begin{figure}[h!]
\begin{center}
\includegraphics[width=0.750\linewidth]{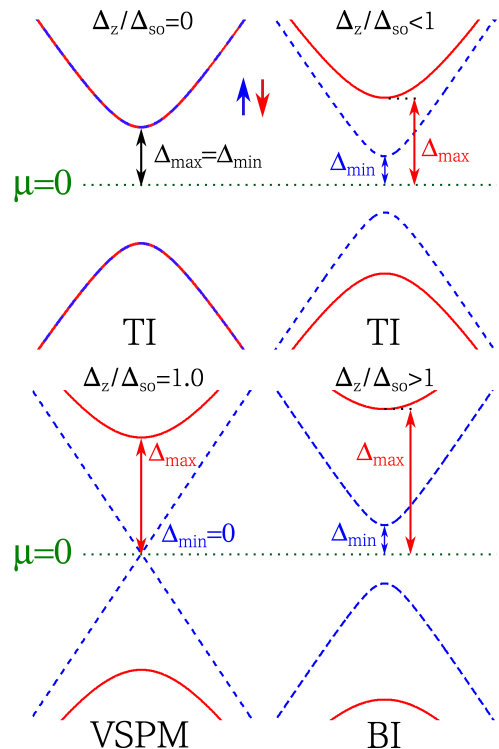}
\end{center}
\caption{\label{fig:Energy}(Color online) Low-energy band structure of a buckled honeycomb lattice at the $K$ point for key values of sublattice potential difference ($\Delta_z$).  Top: (left) With no potential difference, the system is a TI with a band gap of $2\Delta_{\rm so}$. (right)  With a finite $\Delta_z$ such that $\Delta_z<\Delta_{\rm so}$, the system remains a TI but now has two spin split bands the lower(upper) of which decreases(increases) with increased $\Delta_z$.  Bottom: (left) when $\Delta_z=\Delta_{\rm so}$, the lowest gap closes ($\Delta_{\rm min}=0$) and the system is a VSPM. (right) As $\Delta_z$ is increased further such that $\Delta_z>\Delta_{\rm so}$, the gaps reopen and both increase with $\Delta_z$.  The system is now a BI.  In all cases, dashed blue lines correspond to spin up bands and solid red to spin down.  The spin labels switch at $K^\prime$.
}
\end{figure}
Plots of the three regimes (TI, VSPM and BI) are shown in Fig.~\ref{fig:Energy} at the $K$ point.  Here spin up is given by the dashed blue bands and spin down by the solid red bands; the spin labels are reversed at the other valley.

Assuming a tight-binding hopping amplitude of $t=1.6$eV\cite{Liu:2011a}, this low-energy approximation to the band structure holds quite well for energies between approximately $\pm 800$meV\cite{Ezawa:2012a}.

\section{Polarization Function: $\Pi(\omega,q)$}

Many physical properties rely on the dynamical dielectric function $\varepsilon(\omega,q)$.  This function has been studied extensively in a variety of 2D systems beyond the usual electron gas\cite{Stern:1967}. For example, in addition to the works on graphene-related systems previously mentioned, there has been work on semiconducting systems, such as quantum wells, including SOC\cite{Pletyukhov:2006, Badalyan:2009, Badalyan:2010, Kernreiter:2010, Schliemann:2010, Schliemann:2011, Scholz:2013a}.  In the random phase approximation (RPA), the dielectric function is given by
\begin{align}\label{Dielectric}
\varepsilon(\omega,q)=1-V(q)\Pi(\omega,q),
\end{align}
where $V(q)=2\pi\alpha/q$ is the 2D Coulomb potential and $\alpha=e^2/(\hbar\varepsilon_0v)$ is the effective fine structure constant in which $e$ is the elementary charge and $\varepsilon_0$ is the bare dielectric constant.  Therefore, to access information about collective excitations and many-body effects, one needs to calculate the dynamical polarization function $\Pi(\omega,q)$.  In what follows, factors of $\hbar$ and $v$ are ignored.

In the one-loop approximation, the polarization function is found by solving\cite{Mahan:1990,  Gorbar:2002, Wunsch:2006, Hwang:2007, Pyat:2008, Pyat:2009, Rajdeep:2010, Scholz:2012, Scholz:2013}
\begin{align}\label{Pol}
\Pi(\omega,q)=\frac{1}{8\pi^2}\sum_{\sigma,\xi=\pm 1}\int d^2k&\sum_{\lambda,\lambda^\prime=\pm 1}\left(1+\lambda\lambda^\prime\frac{\bm{k}\cdot(\bm{q}+\bm{k})+\Delta_{\sigma\xi}^2}{E_{\bm{k}}E_{\bm{k}+\bm{q}}}\right)\notag\\
&\times\frac{n_F(\lambda E_{\bm{k}})-n_F(\lambda^\prime E_{\bm{k}+\bm{q}})}{\lambda E_{\bm{k}}-\lambda^\prime E_{\bm{k}+\bm{q}}-\omega-i0^+}.
\end{align}
Here, we work at zero temperature so that the Fermi functions $n_F(z)$ can be replaced by step functions.  Since the polarization function depends on the absolute value of the chemical potential ($\mu$) and given the general relation\cite{Pyat:2009} $\Pi(-\omega,q)=\Pi(\omega,q)^*$, we only present the results for $\mu>0$ and $\omega>0$.  Eqn.~\eqref{Pol} can be solved analytically\cite{Pyat:2008, Pyat:2009} to give
\begin{align}\label{PolFull}
\Pi(\omega,q)=\sum_{\sigma,\xi=\pm 1} \bigg[ &\Pi_0^{\sigma\xi}(\omega,q)\Theta(\Delta_{\sigma\xi}-\mu)\notag\\
&+\Pi_1^{\sigma\xi}(\omega,q)\Theta(\mu-\Delta_{\sigma\xi})\bigg] ,
\end{align}
where $\Pi_j^{\sigma\xi}(\omega,q)=\text{Re}\Pi_j^{\sigma\xi}(\omega,q)+i\text{Im}\Pi_j^{\sigma\xi}(\omega,q)$.  

If $\mu<\Delta_{\sigma\xi}$ the imaginary part of the polarization is given by
 \begin{align}\label{ImagPol0}
\text{Im}\Pi_0^{\sigma\xi}(\omega,q)=-\frac{q^2}{16\sqrt{\omega^2-q^2}}\Theta(\omega^2-q^2-4\Delta_{\sigma\xi}^2)\notag\\
\times\left(1+\frac{4\Delta_{\sigma\xi}^2}{\omega^2-q^2}\right)
\end{align} 
and the Kramers-Kronig related real part is
\begin{align}\label{RealPol0}
\text{Re}\Pi_0^{\sigma\xi}(\omega,q)=-\frac{q^2}{4\pi}\bigg\lbrace\frac{\Delta_{\sigma\xi}}{q^2-\omega^2}+\frac{q^2-\omega^2-4\Delta_{\sigma\xi}^2}{4|q^2-\omega^2|^{3/2}}\notag\\
\times\bigg[\Theta(q-\omega)\text{arccos}\frac{q^2-\omega^2-4\Delta_{\sigma\xi}^2}{\omega^2-q^2-4\Delta_{\sigma\xi}^2}\notag\\
-\Theta(\omega-q)\text{ln}\frac{(2\Delta_{\sigma\xi}+\sqrt{\omega^2-q^2})^2}{|\omega^2-q^2-4\Delta_{\sigma\xi}^2|}\bigg]\bigg\rbrace .
\end{align} 

\begin{figure}[h!]
\begin{center}
\includegraphics[width=1.0\linewidth]{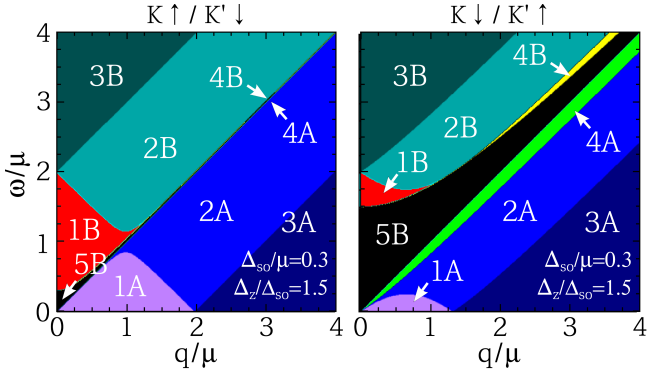}
\end{center}
\caption{\label{fig:Regions}(Color online) Regions for the (left) $K\uparrow/K^\prime\downarrow$ and (right) $K\downarrow/K^\prime\uparrow$ bands in which the polarization function has different expressions. Here, $\Delta_{\rm so}/\mu=0.3$ and $\Delta_z/\Delta_{\rm so}=1.5$.
}
\end{figure}

For $\mu>\Delta_{\sigma\xi}$, the real and imaginary parts are
\begin{align}\label{RealPol}
\text{Re}&\Pi_1^{\sigma\xi}(\omega,q)=-\frac{\mu}{2\pi}+f(\omega,q)\notag\\
&\times\left\lbrace\begin{array}{cc}
0, & \text{1A}\\
G_<\left(\frac{2\mu-\omega}{q}\right), & \text{2A} \\
G_<\left(\frac{2\mu+\omega}{q}\right)+G_<\left(\frac{2\mu-\omega}{q}\right), & \text{3A}\\
G_<\left(\frac{2\mu-\omega}{q}\right)-G_<\left(\frac{2\mu+\omega}{q}\right), & \text{4A}\\
G_>\left(\frac{2\mu+\omega}{q}\right)-G_>\left(\frac{2\mu-\omega}{q}\right), & \text{1B}\\
G_>\left(\frac{2\mu+\omega}{q}\right), & \text{2B}\\
G_>\left(\frac{2\mu+\omega}{q}\right)-G_>\left(\frac{\omega-2\mu}{q}\right), & \text{3B}\\
G_>\left(\frac{\omega-2\mu}{q}\right)+G_>\left(\frac{2\mu+\omega}{q}\right), & \text{4B}\\
G_0\left(\frac{2\mu+\omega}{q}\right)-G_0\left(\frac{2\mu-\omega}{q}\right), & \text{5B}
\end{array}\right.
\end{align}
and
\begin{align}\label{ImagPol}
\text{Im}&\Pi_1^{\sigma\xi}(\omega,q)=-f(\omega,q)\notag\\
&\times\left\lbrace\begin{array}{cc}
G_>\left(\frac{2\mu+\omega}{q}\right)-G_>\left(\frac{2\mu-\omega}{q}\right), & \text{1A}\\
G_>\left(\frac{2\mu+\omega}{q}\right), & \text{2A} \\
0, & \text{3A}\\
0, & \text{4A}\\
0, & \text{1B}\\
-G_<\left(\frac{2\mu-\omega}{q}\right), & \text{2B}\\
\pi(2-x_0^2), & \text{3B}\\
\pi(2-x_0^2), & \text{4B}\\
0, & \text{5B}
\end{array}\right. ,
\end{align}
where we have used the definitions
\begin{align}
f(\omega,q)=\frac{q^2}{16\pi\sqrt{|q^2-\omega^2|}},
\end{align}
\begin{align}
x_0=\sqrt{1+\frac{4\Delta_{\sigma\xi}^2}{q^2-\omega^2}},
\end{align}
\begin{align}
G_<(x)=x\sqrt{x_0^2-x^2}-(2-x_0^2)\text{arccos}(x/x_0),
\end{align}
\begin{align}
G_>(x)=x\sqrt{x^2-x_0^2}-(2-x_0^2)\text{arccosh}(x/x_0),
\end{align}
and
\begin{align}
G_0(x)=x\sqrt{x^2-x_0^2}-(2-x_0^2)\text{arcsinh}\left(x/\sqrt{-x_0^2}\right).
\end{align}
The regions specified in Eqns.~\eqref{RealPol} and \eqref{ImagPol} are given by
\begin{align}\label{regions}
\begin{array}{cc}
\rm{1A}: & \omega<\mu-\sqrt{(q-k_F^{\sigma\xi})^2+\Delta_{\sigma\xi}^2},\\
\rm{2A}: & \pm\mu\mp\sqrt{(q-k_F^{\sigma\xi})^2+\Delta_{\sigma\xi}^2}<\omega\\
 & <-\mu+\sqrt{(q+k_F^{\sigma\xi})^2+\Delta_{\sigma\xi}^2},\\
\rm{3A}: & \omega<-\mu+\sqrt{(q-k_F^{\sigma\xi})^2+\Delta_{\sigma\xi}^2},\\
\rm{4A}: & -\mu+\sqrt{(q+k_F^{\sigma\xi})^2+\Delta_{\sigma\xi}^2}<\omega <q,\\
\rm{1B}: & q<2k_F^{\sigma\xi}, \sqrt{q^2+4\Delta_{\sigma\xi}^2}<\omega\\
 & <\mu+\sqrt{(q-k_F^{\sigma\xi})^2+\Delta_{\sigma\xi}^2},\\
\rm{2B}: & \mu+\sqrt{(q-k_F^{\sigma\xi})^2+\Delta_{\sigma\xi}^2}<\omega\\
 & <\sqrt{(q+k_F^{\sigma\xi})^2+\Delta_{\sigma\xi}^2},\\
\rm{3B}: & \mu+\sqrt{(q+k_F^{\sigma\xi})^2+\Delta_{\sigma\xi}^2}<\omega,\\
\rm{4B}: & q>2k_F^{\sigma\xi}, \sqrt{q^2+4\Delta_{\sigma\xi}^2}<\omega\\
 & <\mu+\sqrt{(q-k_F^{\sigma\xi})^2+\Delta_{\sigma\xi}^2},\\
\rm{5B}: & q<\omega<\sqrt{q^2+4\Delta_{\sigma\xi}^2},
\end{array}
\end{align}
where
\begin{align}
k_F^{\sigma\xi}=\sqrt{\mu^2-\Delta_{\sigma\xi}^2}.
\end{align}
These results reduce to those of Refs.~\cite{Pyat:2008, Pyat:2009} in the limit $\Delta_{\sigma\xi}\rightarrow\Delta$.
The regions given by Eqn.~\eqref{regions} are shown in Fig.~\ref{fig:Regions} for $\Delta_{\rm min}$ (left frame) and $\Delta_{\rm max}$ (right frame) when $\Delta_{\rm so}/\mu=0.3$ and $\Delta_z/\Delta_{\rm so}=1.5$.

\begin{figure}[h!]
\begin{center}
\includegraphics[width=1.0\linewidth]{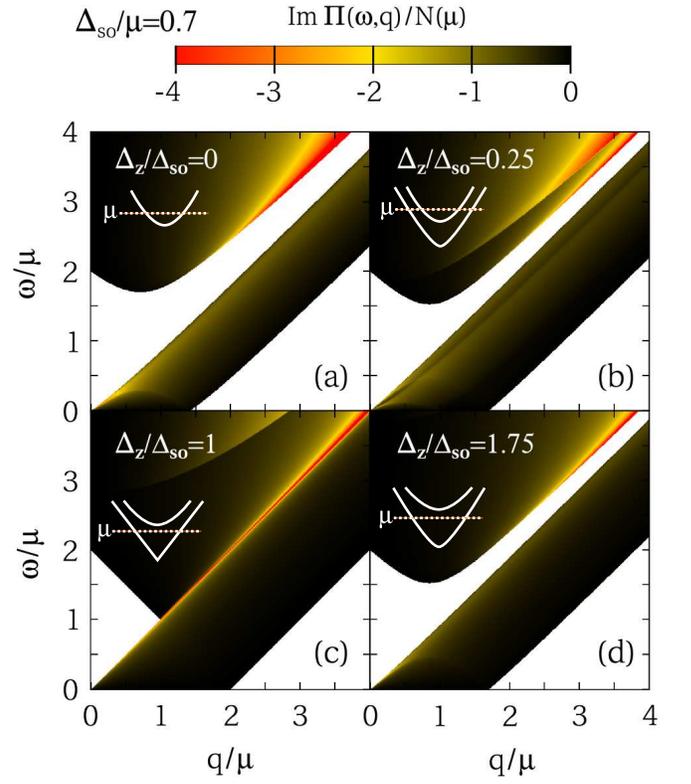}
\end{center}
\caption{\label{fig:ImagPolD07}(Color online) Imaginary part of the polarization function scaled by $N(\mu)=2\mu/\pi$ (density of states of graphene at $\mu$) for $\Delta_{\rm so}/\mu=0.7$ and varying $\Delta_z$. When $\Delta_z=0$ (upper left frame), the Im$\Pi(\omega,q)$ is the familiar case of gapped graphene.  As $\Delta_{z}$ is applied it splits into the sum of two gapped systems with gaps given by $\Delta_{\rm min}$ and $\Delta_{\rm max}$.  When $\Delta_z=\Delta_{\rm so}$ (lower left frame), the polarization function is ungapped and can be understood as the sum of the ungapped graphene system and a system with gap $\Delta_{\rm max}$.  The regions where Im$\Pi(\omega,q)=0$ (white space) correspond to values of $q$ and $\omega$ for which there is no damping of a collective charge oscillation.  Inset: the location of the chemical potential relative to the upper part of the band structure.
}
\end{figure}

Plots of Im$\Pi(\omega,q)$ can be seen in Fig.~\ref{fig:ImagPolD07} for $\Delta_{\rm so}/\mu=0.7$ and varying $\Delta_z$ (i.e the chemical potential is situated above both gaps).
The results for the polarization are essentially the sum of two gapped systems with gaps $\Delta_{\rm min}$ and $\Delta_{\rm max}$.  Thus, as one might expect, when $\Delta_z$ is applied such that $\Delta_z<\Delta_{\rm so}$, the $\omega>q$ regions splits into two with one moving toward the $\omega=q$ line and the other moving away as $\Delta_z$ increases.  When $\Delta_z=\Delta_{\rm so}$, the system is in the VSPM phase and Im$\Pi(\omega,q)$ is not gapped.  As $\Delta_z$ is increased further, the two gaps in the $\omega>q$ region grow with $\Delta_z$. Examining when these signatures of the three insulating regimes occur, may allow for an experimental determination of $\Delta_{\rm so}$.  The imaginary part of the polarization function plays an important role in determining the behaviour of the plasmons.  The regions in $(q,\omega)$ space of non-zero polarization correspond to regions in which collective oscillations are damped.  This is discussed further in Sec.~IV.

The optical conductivity is related to the polarization function through\cite{Bruus:2005,Wunsch:2006, Scholz:2011, LeBlanc:2012} $\sigma(\omega)=\lim_{q\rightarrow 0} ie^2\omega\Pi(\omega,q)/q^2$.  Therefore, the behaviour of Im$\Pi(\omega,q)$ along the $q=0$ line is directly related to the absorptive part of the optical conductivity which can be seen in Ref.~\cite{Stille:2012}.  No spin and valley polarization information will be apparent without the use of polarized light\cite{Stille:2012} or magnetic fields\cite{Tabert:2013a, Tabert:2013c}.  With the expressions given below, one may form the quantity $\sigma(\omega,q)$ in the limit of the impurity scattering rate going to zero from $\sigma(\omega, q)= ie^2\omega\Pi(\omega,q)/q^2$ which has been important for the recent developments in infrared nanoscopy\cite{Fei:2011,LeBlanc:2012, Ashby:2012}.  A discussion of deviations from this simple formula due to impurity scattering is given in Ref.~\cite{Ashby:2012}.

\begin{figure}[h!]
\begin{center}
\includegraphics[width=1.0\linewidth]{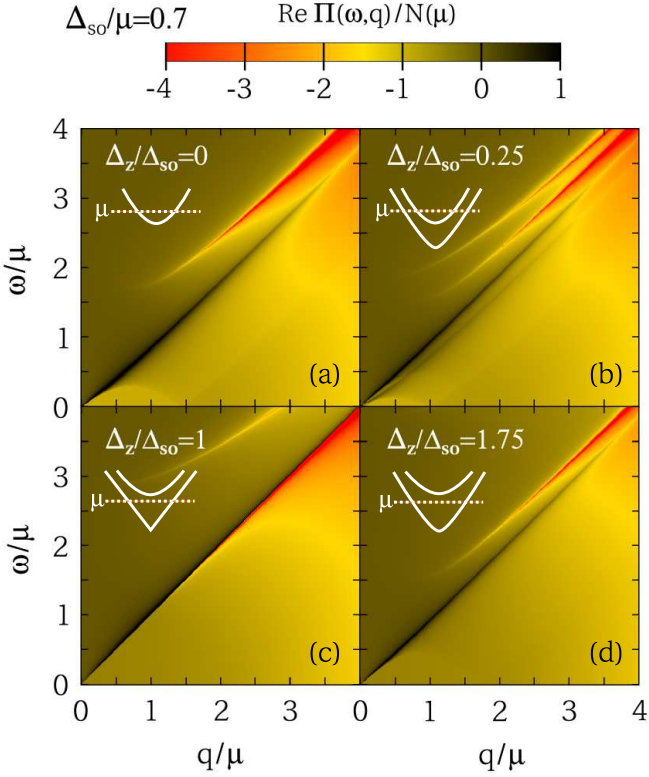}
\end{center}
\caption{\label{fig:RealPolD07}(Color online) Real part of the polarization function for $\Delta_{\rm so}/\mu=0.7$ and $\Delta_z$ chosen to correspond to Fig.~\ref{fig:ImagPolD07}.  Again, it behaves as the sum of two gapped systems.  Inset: the location of the chemical potential relative to the gap(s).
}
\end{figure}

\begin{figure}[h!]
\begin{center}
\includegraphics[width=1.0\linewidth]{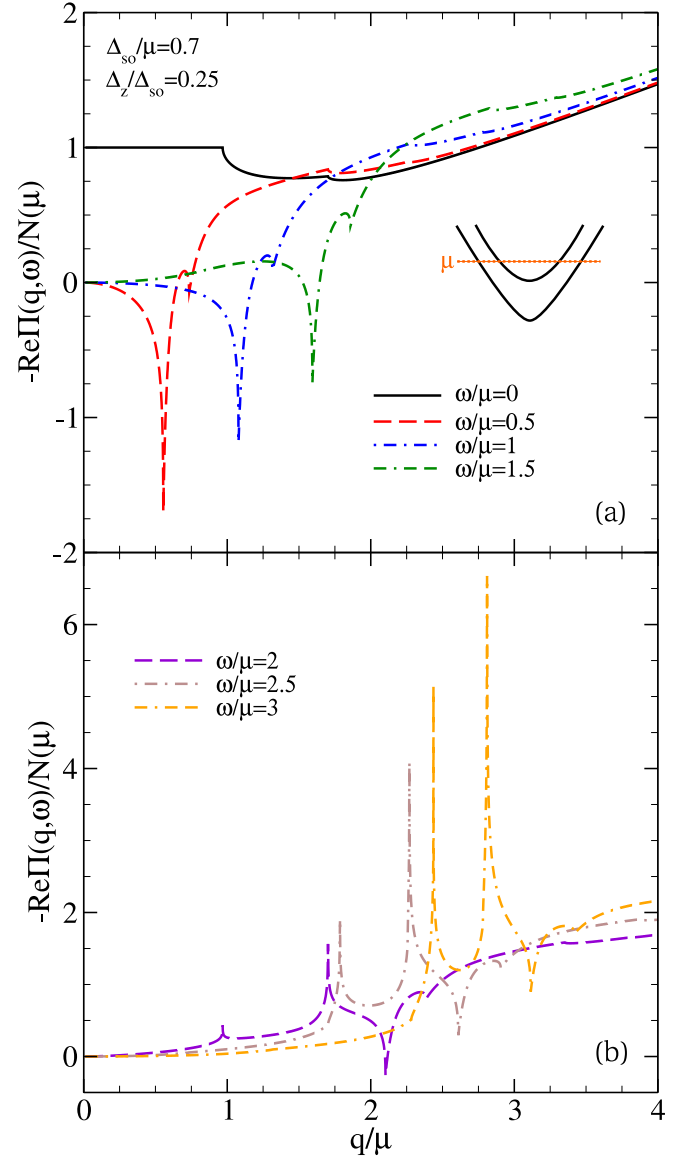}
\end{center}
\caption{\label{fig:RealPol-Wcut}(Color online) Constant frequency cuts of the real part of the polarization function as a function of $q$ for $\Delta_{\rm so}/\mu=0.7$ and $\Delta_z/\Delta_{\rm so}=0.25$ [see Fig.~\ref{fig:RealPolD07}(b)]. (a) $\omega/\mu=0, 0.5, 1$ and 1.5. (b) $\omega/\mu=2,2.5$ and 3. Inset: the relative location of $\mu$ with respect to the band structure.
}
\end{figure}

Re$\Pi(\omega,q)$ is plotted in Fig.~\ref{fig:RealPolD07} with $\Delta_{\rm so}/\mu$ and $\Delta_z$ chosen to correspond with Fig.~\ref{fig:ImagPolD07}.
Several frequency cuts of Fig.~\ref{fig:RealPolD07}(b) are shown in Fig.~\ref{fig:RealPol-Wcut}.  
While the imaginary part of the polarization determines the damping of the plasmon, the real part of the dynamical polarization comes in to determining the location of the plasmon branch in $(q,\omega)$ space.  Again, Sec.~IV contains a further discussion on the plasmon dispersion. 

\begin{figure}[h!]
\begin{center}
\includegraphics[width=1.0\linewidth]{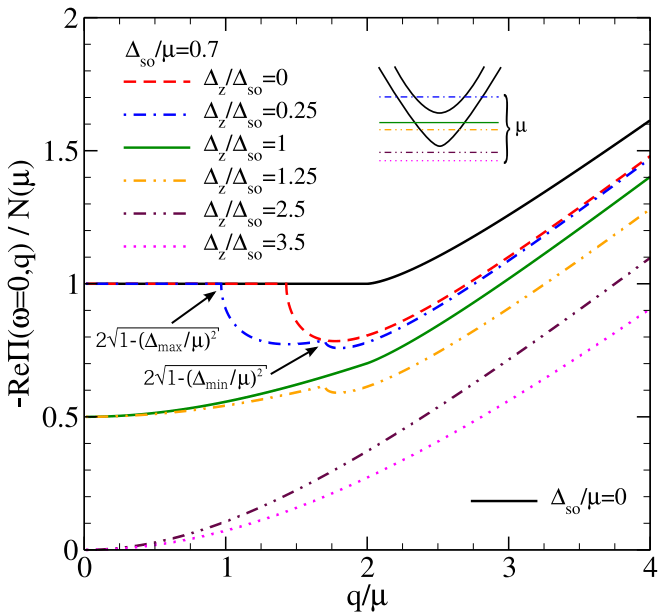}
\end{center}
\caption{\label{fig:Pol-W0}(Color online) Static polarization scaled by $N(\mu)$ for $\Delta_{\rm so}/\mu=0.7$ and varying $\Delta_z$.  The graphene result ($\Delta_{\rm so}=0$) is given by the solid black curve for comparison.  Inset: relative location of the chemical potential for finite $\Delta_z$ (not drawn to scale) [see text for a further discussion].  Like graphene, when the chemical potential is above both gaps (dashed red and dash-dotted blue curves), the polarization is constant.  At $q/\mu=2\sqrt{1-(\Delta_{\rm max}/\mu)^2}$ and $q/\mu=2\sqrt{1-(\Delta_{\rm min}/\mu)^2}$, there are dips in the polarization after which the graphene-like interband behaviour is retained.  When $\Delta_{\rm min}<\mu<\Delta_{\rm max}$ so $\mu$ lies between the two gaps (solid green and dash-double-dotted orange curves), the $q=0$ polarization is half that of graphene and never remains constant.  There is a small dip at $q/\mu=2\sqrt{1-(\Delta_{\rm min}/\mu)^2}$ before a continual increase is observed.  When $\mu$ sits in both gaps (dash-double-dotted purple and dotted magenta curves), $\Pi(0,0)=0$ and there are no cusps at higher $q$.  For all cases, the polarization function at a given $q$ is always less than or equal to that of a lower $\Delta_z$ value.
}
\end{figure}

The static limit of Re$\Pi(\omega,q)$ is of particular importance as it determines the screened potential of a charge impurity.  In the limit $\omega\rightarrow 0$, Im$\Pi(\omega,q)\rightarrow 0$ and the static polarization is given by
\begin{align}\label{Lindhard}
\Pi(0,q)=-\frac{\mu}{2\pi}\sum_{\sigma,\xi=\pm 1}\left[F(q)\Theta(\Delta_{\sigma\xi}-\mu)+G(q)\Theta(\mu-\Delta_{\sigma\xi})\right]
\end{align} 
where\cite{Pyat:2009}
\begin{align}
F(q)=\frac{\Delta_{\sigma\xi}}{2\mu}+\frac{q^2-4\Delta_{\sigma\xi}^2}{4q\mu}\text{arcsin}\sqrt{\frac{q^2}{q^2+4\Delta_{\sigma\xi}^2}},
\end{align} 
and
\begin{align}
G(q)=1-\Theta(q-2k_F^{\sigma\xi})\left[\frac{\sqrt{q^2-4(k_F^{\sigma\xi})^2}}{2q}-\frac{q^2-4\Delta_{\sigma\xi}^2}{4q\mu}\right.\notag\\
\left.\times\text{arctan}\left(\frac{\sqrt{q^2-4(k_F^{\sigma\xi})^2}}{2\mu}\right)\right].
\end{align}  
A plot of $\Pi(0,q)$ for $\Delta_{\rm so}/\mu=0.7$ and varying $\Delta_z$ is shown in Fig.~\ref{fig:Pol-W0}.
The graphene case ($\Delta_{\rm so}/\mu=0$) is shown for comparison (solid black curve).  The static polarization of graphene is well known\cite{Shung:1986, Wunsch:2006, Hwang:2007, Gonzalez:1994, Roldan:2013, Yuan:2013}.  A single layer of graphene displays a linear contribution to the static polarization function of $\Pi(0,q)=-N(\mu)\pi q/(8k_F)$.  This is associated with interband transitions.  There is also an intraband contribution of the form $\Pi(0,q)= -N(\mu)[1-\pi q/(8k_F)]$ for $q\leq 2k_F$ and $\Pi(0,q)=-N(\mu)[1-(1/2)\sqrt{1-4k_F^2/q^2}-(q/4k_F){\rm sin^{-1}}(2k_F/q)]$ for $q>2k_F$.  Combining the two results gives a constant polarization of $-N(\mu)$ for $q\leq 2k_F$ which is typical of metallic screening\cite{Fetter:2003, Yuan:2013}.  For large $q$, the interband term results in a linear contribution which signifies insulating-like screening\cite{Yuan:2013}.  Returning to Fig.~\ref{fig:Pol-W0}, the dashed red and dashed-dotted blue curve correspond to $\mu>\Delta_{\rm max}$ such that the chemical potential is above the gap(s).  Like graphene and the two-dimensional electron gas, for $q/\mu<2\sqrt{1-(\Delta_{\rm max}/\mu)^2}$, the polarization is constant.  After this value, there is a decrease in the magnitude of the polarization function before it begins to increase linearly due to the interband transitions.  The solid green and dash-double-dotted orange curves correspond to $\Delta_{\rm min}<\mu<\Delta_{\rm max}$ so $\mu$ lies between the two gaps.  In this case, the magnitude of the polarization begins at half the graphene value and shows a general increasing trend with a small dip at $q/\mu=2\sqrt{1-(\Delta_{\rm min}/\mu)^2}$.  Finally, the dash-double-dotted purple and dotted magenta curves result from the chemical potential sitting in both gaps.  In this case, $\Pi(0,0)= 0$ and $\Pi(0,q\rightarrow 0)\sim-\sum_{\sigma\xi}q^2/(12\pi\Delta_{\sigma\xi})$.  In all cases, as $\Delta_z$ is increased, the magnitude of the polarization at a given $q$ is less than or equal to that of lower $\Delta_z$.  When using the inset of Fig.~\ref{fig:Pol-W0}, it is important to note that the chemical potential is fixed and it is the band structure that is varying.  The inset merely shows where the chemical potential sits relative to the two bands.

\section{Loss Function and Plasmons}

\begin{figure}[h!]
\begin{center}
\includegraphics[width=1.0\linewidth]{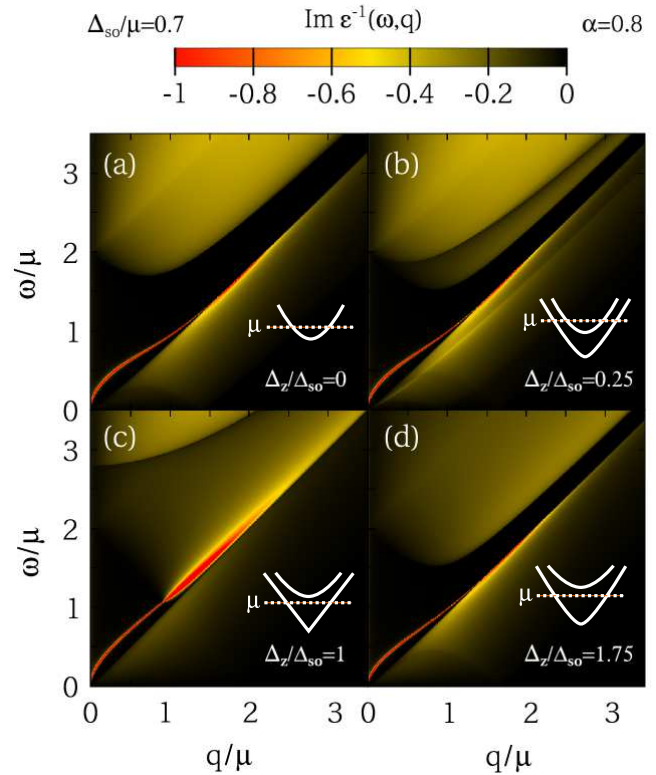}
\end{center}
\caption{\label{fig:Loss}(Color online) Energy loss function for $\alpha=0.8$, $\Delta_{\rm so}/\mu=0.7$ and varying $\Delta_z$ chosen to correspond with Fig.~\ref{fig:ImagPolD07}.  The red branch signifies the plasmon excitation.  The plasmon is undamped in the regions that correspond to Im$\Pi(\omega,q)=0$ (see Fig.~\ref{fig:ImagPolD07}).  When the imaginary part of the polarization function is nonzero, the plasmon is quickly damped out in $(q,\omega)$ space. Due to the $1/[1-V(q)\Pi(\omega,q)]$ dependence, the energy loss function is not simply the sum of two gapped systems.  Inset: the location of the chemical potential relative to the band structure.
}
\end{figure}

\begin{figure}[h!]
\begin{center}
\includegraphics[width=1.0\linewidth]{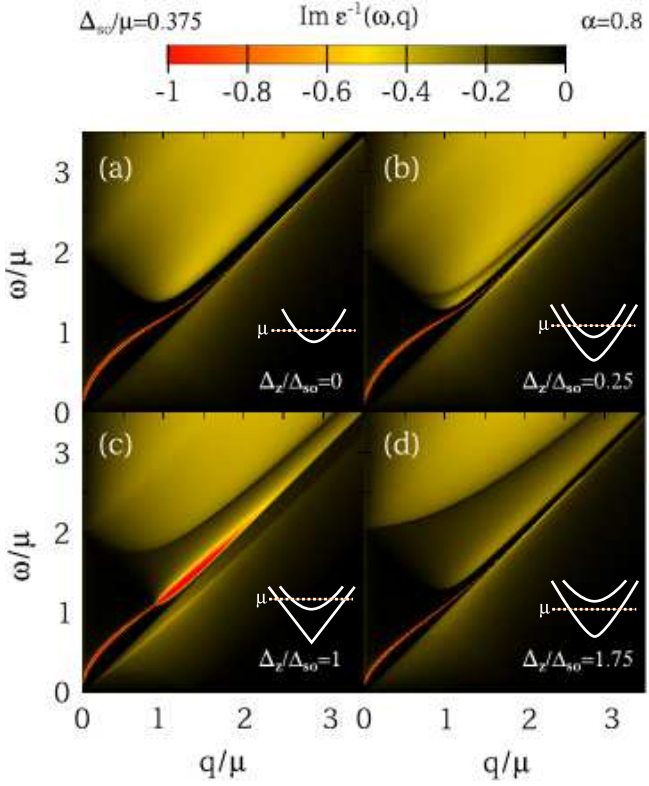}
\end{center}
\caption{\label{fig:Loss-0375}(Color online) Energy loss function for $\alpha=0.8$, $\Delta_{\rm so}/\mu=0.375$ and varying $\Delta_z$.  Inset: the location of the chemical potential relative to the band structure.
}
\end{figure}

In the RPA, the plasmon dispersion $\omega_p(q)$ is given by the poles of the loss function Im$\varepsilon^{-1}(\omega,q)$.  This function is plotted in Fig.~\ref{fig:Loss} for $\alpha=0.8$, $\Delta_{\rm so}/\mu=0.7$ and $\Delta_z$ set to match Figs.~\ref{fig:ImagPolD07} and \ref{fig:RealPolD07}.  
For graphene, the fine structure constant has been experimentally determined for a variety of substrates\cite{Hwang:2012}.  On SiC, it is reported as $\alpha\approx 4.4$, for $h$-BN $\alpha\approx 7.7$ and for Quartz $\alpha\approx 17.7$.  The value of $\alpha$ used herein was chosen to elucidate interesting features on a tractable energy scale and not to mimic an experimental value.  Another set of plots of Im$\varepsilon^{-1}(\omega,q)$ is shown in Fig.~\ref{fig:Loss-0375} for $\alpha=0.8$, $\Delta_{\rm so}/\mu=0.375$ and varying $\Delta_z$.  This set of parameters was chosen to yield multiple solutions to the approximate equation for determining the plasmon frequency.  This is discussed in more detail at the end of this section.
In each case, the red branch corresponds to the pole of Im$\varepsilon^{-1}(\omega,q)$ and thus describes the plasmon excitation.  In the $(q,\omega)$ domain for which Im$\Pi(\omega,q)=0$ (white regions of Fig.~\ref{fig:ImagPolD07}), the plasmon is undamped; however, when the imaginary part of the polarization is non-zero, collective oscillations are quickly damped.  As the polarization enters in the denominator of the loss function (i.e.  $\varepsilon^{-1}\propto 1/[1-V(q)\Pi(\omega,q)])$, the plasmon dispersion is not simply the sum of two gapped systems.

A similar situation of spin-split bands is found in MoS$_2$ (see Ref.~\cite{Scholz:2013}) although the details are quite different.  Due to the large band gap in the MoS$_2$ band structure, the plasmons damp out in the intraband portion of the particle-hole continuum\cite{Scholz:2013} (PHC) instead of the interband damping seen in graphene.  In silicene, intraband damping is observed for most values of $\Delta_z$; however, for $\Delta_z\approx\Delta_{\rm so}$, the plasmon branch damps out in the interband region of the PHC due to the closing of the $\Delta_{\rm min}$ gap.  In addition, as our model Hamiltonian is particle-hole symmetric, there is no difference for the plasmon dispersion curve for the particles and holes as found for MoS$_2$. 

\begin{figure}[h!]
\begin{center}
\includegraphics[width=1.0\linewidth]{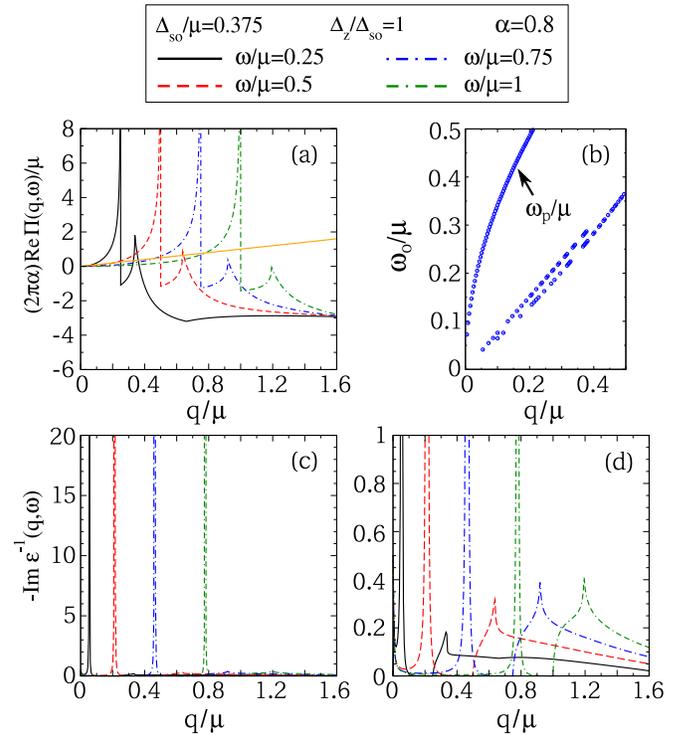}
\end{center}
\caption{\label{fig:PlasNum-3Branch}(Color online) (a) Frequency cuts of $2\pi\alpha$Re$\Pi(\omega,q)/\mu$ for $\alpha=0.8$, $\Delta_{\rm so}/\mu=0.375$ and $\Delta_z/\Delta_{\rm so}=1$.  The intersection of $2\pi\alpha$Re$\Pi(\omega,q)/\mu$ and the line of unit slope (solid orange line) represent solutions to Eqn.~\eqref{PlasDef}.  (b) Plasmon frequency as a function of $q$ corresponding to the solutions of Eqn.~\eqref{PlasDef}. (c) Frequency cuts of the loss function in which we can see that there is one pole per frequency.  (d) Zoomed in version of (c).
}
\end{figure}

The plasmon branch can also be obtained by finding the zeros of the dielectric function $\varepsilon(\omega_p-i\gamma,q)$, where $\gamma$ is the plasmon decay rate.  For weak damping, it is sufficient to solve\cite{Fetter:2003, Wunsch:2006, Pyat:2008, Pyat:2009, Scholz:2012}, 
\begin{align}\label{PlasDef}
q-2\pi\alpha\text{Re}\Pi(\omega,q)=0,
\end{align}
where the decay rate of the plasmon is determined by
\begin{align}
\gamma=\frac{\rm{Im}\Pi(\omega_p,q)}{(\partial/\partial\omega)\rm{Re}\Pi(\omega,q)|_{\omega=\omega_p}}.
\end{align}
As we are primarily interested in the low $q$ behaviour of the plasmon which exists when Im$\Pi(\omega,q)=0$ and thus $\gamma=0$, we only report the results of Eqn.~\eqref{PlasDef}.  Several slices of $2\pi\alpha$Re$\Pi(\omega,q)/\mu$ are shown in Fig.~\ref{fig:PlasNum-3Branch}(a) for $\alpha=0.8$, $\Delta_{\rm so}/\mu=0.375$ and $\Delta_z/\Delta_{\rm so}=1$.  The solutions of Eqn.~\eqref{PlasDef} are given by the intersection of $2\pi\alpha$Re$\Pi(\omega,q)/\mu$ versus $q/\mu$ with the line of unit slope\cite{LeBlanc:2014} (solid orange line).  We can see that for low $\omega$ (first two curves), the function intersects the line three times and, therefore, there are three solutions at those frequencies ($\omega_o$).  This is shown in Fig.~\ref{fig:PlasNum-3Branch}(b).  The two larger $q$ contributions are not physical\cite{Pyat:2009, Scholz:2012} as Eqn.~\eqref{PlasDef} is only exact when Im$\Pi(\omega,q)=0$ and these values fall in the region when this is finite.  We can see this by examining cuts of the loss function shown in Fig.~\ref{fig:PlasNum-3Branch}(c) and (d) where we note that for a given $\omega$, there is only one pole and thus only one plasmon branch.  The physical solution is denoted $\omega_p$.

\begin{figure}[h!]
\begin{center}
\includegraphics[width=1.0\linewidth]{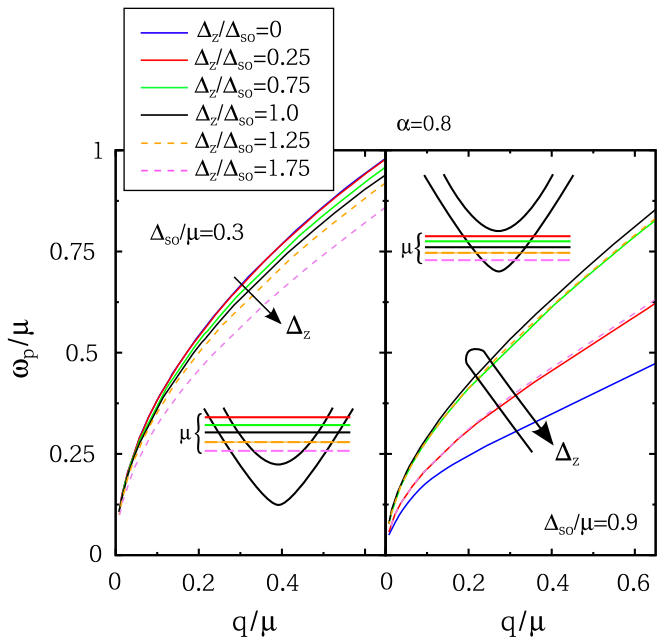}
\end{center}
\caption{\label{fig:PlasNum}(Color online) Low-energy plasmon dispersion for $\alpha=0.8$, varying $\Delta_z$ and (left) $\Delta_{\rm so}/\mu=0.3$ and (right) $\Delta_{\rm so}/\mu=0.9$. Left: For $\Delta_{\rm so}/\mu=0.3$, the chemical potential is greater than $\Delta_{\rm max}$ for all values of $\Delta_z$ shown here.  As $\Delta_z$ is increased, the plasmon branch at a given $q$ decrease in frequency.  Right: For $\Delta_{\rm so}/\mu=0.9$, the nonzero $\Delta_z$ values correspond to $\Delta_{\rm min}<\mu<\Delta_{\rm max}$.  Here the plasmon branch increases with $\Delta_z$ in the TI regime, reaches a maximum in the VSPM phase and then decreases with increased $\Delta_z$ in the BI regime.  Thus, the location of the plasmon branch in $(q,\omega)$ is highly dependent on the sublattice potential difference. Inset: the location of the band structure relative to the chemical potential (not drawn to scale).  Note that the chemical potential is fixed and it is the band structure which is varying in energy.
}
\end{figure}

Using Eqn.~\eqref{PolFull}, the numerical solutions to Eqn.~\eqref{PlasDef} are obtained and are shown in the left and right frames of Fig.~\ref{fig:PlasNum} for $\Delta_{\rm so}/\mu=0.3$ and 0.9, respectively. 
For $\Delta_{\rm so}/\mu=0.3$, the chemical potential is above both gaps for all chosen values of $\Delta_z$.  In this regime, the frequency of the plasmon branch decreases for a given $q$ as the onsite potential difference is increased.  For $\Delta_{\rm so}/\mu=0.9$, the non-zero $\Delta_z$ values result in $\Delta_{\rm min}<\mu<\Delta_{\rm max}$.  Here, the plasmon frequency increase with $\Delta_z$ in the TI regime (while the lower band gap is decreasing) until it reaches a maximum in the VSPM phase and then decreases with increased $\Delta_z$ in the BI regime (while both band gaps are increasing with increased $\Delta_z$).  Note, once $\mu<\Delta_{\rm min}$, collective oscillations are not present.  If the sublattice potential difference could be tuned without substantially changing properties of the system (such as electron-electron interactions), the location of the plasmon branch could be manipulated.  This may be of interest in the field of plasmonics.   In addition, by varying $\Delta_z$ at fixed $\Delta_{\rm min}<\mu<\Delta_{\rm max}$, the change in behaviour of the plasmon branch could allow for an experimental determination of $\Delta_{\rm so}$.

It is possible to capture the behaviour of the plasmon analytically in the limit $q\ll\omega\ll\mu$.  If $\mu>\Delta_{\rm{max}}$, the small $q$ and $\omega$ behaviour is
\begin{align}\label{Plasmon2Gap}
\frac{\omega_p(q)}{\mu}\approx\sqrt{\frac{\alpha q}{2\mu}\sum_{\sigma,\xi=\pm 1}\left[1-\left(\frac{\Delta_{\sigma\xi}}{\mu}\right)^2\right]},
\end{align} %pull out 4 from alpha! so all my worked out formulas (in my notes) are off by 4
or, equivalently,
\begin{align}
\frac{\omega_p(q)}{\mu}\approx\sqrt{\frac{\alpha q}{\mu}\left[2-\left(\frac{\Delta_{{\rm max}}}{\mu}\right)^2-\left(\frac{\Delta_{{\rm min}}}{\mu}\right)^2\right]}.
\end{align}
In terms of the individual Fermi momenta of the two spin split bands, $\omega_p(q)$ can be written as
\begin{align}
\frac{\omega_p(q)}{\mu}\approx\sqrt{\frac{\alpha q}{\mu^3}\left[\left(k_F^{\rm min}\right)^2+\left(k_F^{\rm max}\right)^2\right]},
\end{align}
where $k_F^{\rm max}=\sqrt{\mu^2-\Delta_{\rm max}^2}$ and $k_F^{\rm min}=\sqrt{\mu^2-\Delta_{\rm min}^2}$.  Thus, the plasmon dispersion is dependent on the Fermi momentum of the individual spin-split bands.

If $\Delta_{\rm{min}}<\mu<\Delta_{\rm{max}}$,
\begin{align}\label{Plasmon1Gap}
\frac{\omega_p(q)}{\mu}\approx\sqrt{\frac{\alpha q}{\mu}\left(1-\frac{\Delta_{\rm min}^2}{\mu^2}\right)}.
\end{align}

If the chemical potential lies in both gaps, there is no plasmon branch.
 
\section{Screening of a Charged Impurity}

In the RPA, the static potential of a screened impurity of charge $Q$ is given by\cite{Stern:1967,Gamayun:2011,Scholz:2012,Scholz:2013}
\begin{align}\label{ScreenedPotential}
\Phi(r)=\frac{Q}{\varepsilon_0}\int_0^\infty dq\frac{J_0(qr)}{\varepsilon(0,q)},
\end{align}
where $J_0(x)$ is the zeroth Bessel function of the first kind.  The Lighthill theorem\cite{Lighthill:1958, Scholz:2013} implies that the non-analytic points of the dielectric function yield the asymptotic behaviour of the screening potential.  Examining Fig.~\ref{fig:Pol-W0}, we see that singular points arise in the derivative of the static polarization when $q=2k_F^{\rm max}$ and $2k_F^{\rm min}$.  Thus, sinusoidal Friedel oscillations are observed in $\Phi(r)$\cite{Scholz:2013}.

\begin{figure}[h!]
\begin{center}
\includegraphics[width=1.0\linewidth]{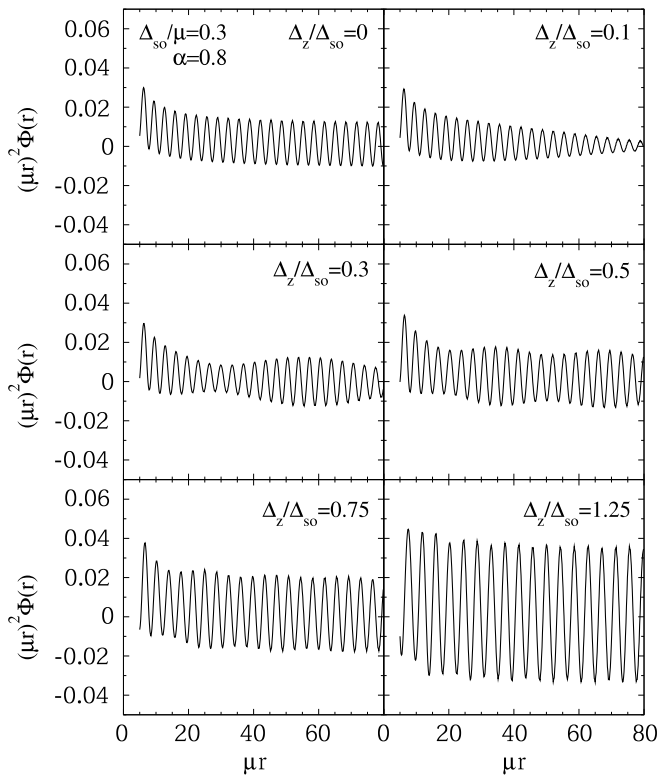}
\end{center}
\caption{\label{fig:Screened}(Color online) Screened potential of a charged impurity in units of $Q/\varepsilon_0$ for $\alpha=0.8$, $\Delta_{\rm so}$ and varying $\Delta_z$.  As a function of $r$, there are two primary decay rates: $1/r^3$ and $1/r^2$ for small and large $r$, respectively.  With the inclusion of $\Delta_z$, a beating effect is observed with the distance between beats decreasing with increasing $\Delta_z$.
}
\end{figure}

Equation~\eqref{ScreenedPotential} can be evaluated numerically where we note that the value of $\Phi(r)$ is primarily determined by the long-wavelength (small $q$) behaviour of the static polarization function [Eqn.~\eqref{Lindhard}].  The numerical evaluation of Eqn.~\eqref{ScreenedPotential} is shown in Fig.~\ref{fig:Screened} for $\alpha=0.8$, $\Delta_{\rm so}/\mu=0.3$ and varying $\Delta_z$.
For $\Delta_z=0$ (upper left frame), we obtain the results of gapped graphene\cite{Pyat:2009, Scholz:2012}.  That is, we have an oscillatory function that decays as $1/r^3$ for small $r$ and $1/r^2$ for large $r$.  When a finite sublattice potential difference is included, a beating effect is observed in the Friedel oscillations for $\mu>\Delta_{\rm max}$. The characteristic length between beats changes as a  function of $\Delta_z$.  As a result, for small $\Delta_z$, which corresponds to the two gaps being close together in energy, the characteristic wavelength of the beats is long.  As $\Delta_z$ is increased, the separation between beats decreases due to the difference in the two $k_F$ values becoming large.  For sufficiently large $\Delta_z$, the beating becomes hard to discern.  This effect is only observed if $\mu$ is above both gaps.

Beating of the Friedel oscillations is also seen in other systems with split bands, such as MoS$_2$\cite{Scholz:2013}.  This also results from the two $k_F$ values associated with the two split bands which yield kinks in the static polarization function.  While occupation of the two bands is required here (i.e. $\mu>\Delta_{\rm max}$), our work differs from the MoS$_2$ case as we are interested in examining how the $\Delta_z$ dependence of the $k_F$'s affects the beating phenomenon.

\section{Conclusions}

We apply the dynamic polarization results of gapped graphene\cite{Pyat:2008,Pyat:2009} to a 2D Kane-Mele type topological insulator presenting both numerical and analytical results.  Particular attention is given to a buckled-honeycomb structure where a valley-spin-polarized band structure is attained by creating an onsite potential difference between the $A$ and $B$ sublattices.  We show that while the polarization function of this two-gap system is the sum of two single gapped systems, that is not the case for the plasmon dispersion.  In particular, the imaginary part of $\Pi(\omega,q)$ displays signatures of the three insulating regimes of the system (TI, VSPM and BI) and may be used to elucidate information on the strength of the intrinsic spin-orbit gap of the system.  We show that the static polarization [$\Pi(\omega=0,q)$] can display three distinct behaviours, depending on where the chemical potential falls relative to the gaps, with $q\rightarrow 0$ limits of $N(\mu)$, $N(\mu)/2$ and 0, where $N(\mu)$ is the density of states of graphene.

An undamped plasmon branch is seen for low $q$ and $\omega$.  We examine its behaviour for varying sublattice potential difference with particular attention to the effect of where $\mu$ is placed relative to the two gaps.  We capture the low $q$ and $\omega$ behaviour with analytic formulae.  Indeed, we find that the location of the plasmon in $(q,\omega)$ space is highly dependent on the values of $\Delta_z$ and $\mu$ relative to $\Delta_{\rm so}$.  

We present the numerical results for the effective potential of a statically screened charged impurity.  We observe an overall beating in the Friedel oscillations which results from two singularities in the first derivative of $\Pi(0,q)$ due to the presence of two gaps.  The dependence of this beating phenomenon on the sublattice potential difference is discussed.

The results of this paper apply to buckled honeycomb systems which display a sizable spin-orbit interaction such as silicene and germanene.  We trust that as such samples become more readily available, this work will help guide the experimental community in characterizing the systems and observing many of their exciting and novel properties.

\emph{Note added.}  After submitting this manuscript, a recent preprint appeared\cite{Chang:2014} which also examines similar topics, such as, Friedel oscillations in silicene and confirms the phenomena discussed herein.

\begin{acknowledgments}
This work has been supported by the Natural Science and Engineering Research Council of Canada.  We thank J. P. Carbotte and J. P. F. LeBlanc for discussions.
\end{acknowledgments}

\bibliographystyle{apsrev4-1}
\bibliography{rpa}

\end{document}